\documentclass[11pt,a4paper]{article}
\usepackage{jheppub_kim}
\usepackage{rotating} 
\usepackage{graphicx,epsfig}
\usepackage{amsmath}
\usepackage {amssymb}
\usepackage{subfigure}

\usepackage{relsize}

\providecommand{\U}[1]{\protect\rule{.1in}{.1in}}

\newcommand{\newc}{\newcommand}

\newc{\be}{\begin{equation}}
\newc{\ee}{\end{equation}}

\newc{\ba}{\begin{eqnarray}}
\newc{\ea}{\end{eqnarray}}
\newc{\bea}{\begin{eqnarray*}}
\newc{\eea}{\end{eqnarray*}}
\newc{\D}{\partial}
\newc{\ie}{{\it i.e.} }
\newc{\eg}{{\it e.g.} }
\newc{\etc}{{\it etc.} }
\newc{\etal}{{\it et al.}}
\newc{\lcdm}{$\Lambda$CDM }

\newc{\ra}{\Rightarrow}

\title{Alleviating both $H_0$ and $\sigma_8$ tensions in Tsallis cosmology}
 \author[a,b,c]{Spyros Basilakos}
 \author[d]{Andreas Lymperis}
 \author[e]{Maria Petronikolou}
 \author[a,f,g]{Emmanuel N. Saridakis}

\affiliation[a]{National Observatory of Athens, Lofos Nymfon, 11852 Athens, 
Greece}
\affiliation[b]{Academy of Athens, Research Center for Astronomy and Applied 
Mathematics, Soranou Efesiou 4, 11527, Athens, Greece}

\affiliation[c]{School of Sciences, European University Cyprus, Diogenes 
Street, Engomi, 1516 Nicosia, Cyprus}

\affiliation[d]{Department of Physics, University of Patras, 26500 Patras, 
Greece}

\affiliation[e]{Department of Physics, National Technical University of Athens, 
Zografou Campus GR 157 73, Athens, Greece}

\affiliation[f]{Department of Astronomy, School of Physical Sciences, 
University of Science and Technology of China, Hefei 230026, P.R. China}

 \affiliation[g]{Departamento de Matem\'{a}ticas, Universidad Cat\'{o}lica del 
Norte, 
Avda.
Angamos 0610, Casilla 1280 Antofagasta, Chile}

   \emailAdd{svasil@academyofathens.gr}
\emailAdd{alymperis@upatras.gr}
\emailAdd{petronikoloumaria@mail.ntua.gr}
\emailAdd{msaridak@noa.gr}

\abstract{  
We present how Tsallis cosmology can alleviate both   $H_0$ and $\sigma_8$  
tensions simultaneously. Such a modified cosmological scenario is obtained by 
the application of the gravity-thermodynamics conjecture, but using the 
non-additive Tsallis entropy, instead of the standard Bekenstein-Hawking one. 
Hence, one obtains modified Friedmann equations, with extra terms that depend 
on the new Tsallis exponent $\delta$ that quantifies the departure from 
standard entropy. We  show that for particular $\delta$ choices    we can 
obtain a phantom effective dark energy, which is known to be one of the 
sufficient mechanisms that can alleviate  $H_0$ tension. Additionally, 
for the same parameter choice we obtain  an increased friction term and 
an effective Newton's constant smaller than the usual one, and thus the 
$\sigma_8$  
tension is also solved.
These features 
act as a significant advantage of Tsallis modified cosmology.
}


\begin{document}
\maketitle

\section{Introduction}

The Standard Paradigm of Cosmology, namely $\Lambda$CDM concordance model, has 
been proven very successful in describing the universe evolution, at both 
background and perturbative levels. Nevertheless,  it  
exhibits possible disadvantages, either theoretical or observational  
\cite{Perivolaropoulos:2021jda}. In the first class of potential issues one has 
the non-renormalizability of general 
relativity or the cosmological constant problem. In the second class one may 
find the dynamical behavior of dark energy, the realization of the inflationary 
phase, as well as     various cosmological tensions.

Among cosmological tensions, one has the $H_0$ one, namely the fact that the 
present value of the Hubble parameter is estimated by    
Planck collaboration   to be $H_0 = (67.27\pm 0.60)$ km/s/Mpc 
\cite{Planck:2018vyg}, whereas local  measurements of the $2019$ 
SH$0$ES collaboration ($R19$) lead to $H_0 = (74.03\pm1.42)$ km/s/Mpc. 
Additionally, 
one may have the  $\sigma_8$ tension, which is  related to the matter 
clustering, and the fact that the Cosmic Microwave 
Background (CMB) estimation \cite{Planck:2018vyg} differs from  the 
SDSS/BOSS direct measurement
\cite{Zarrouk:2018vwy,BOSS:2016wmc}.

The first local ladder measurement of 
  $H_0$, obtained using the Hubble Space Telescope (HST) observations of 
Cepheids and SNIa, was $72 \pm 8 $ km/s/Mpc  \cite{HST:2000azd}. Subsequently, 
dedicated efforts were made to enhance the precision of the measurements, 
resulting in a refined estimate of $H_0 = 74.3 \pm 2.2  $  km/s/Mpc 
\cite{Freedman:2012ny}. Additionally, in $2005$, the SH$0$ES Project started and 
proceeded to advance this approach. Over the years, numerous $H_0$ estimations 
have been 
obtained by the SH$0$ES collaboration, with the most recent and refined result 
yielding $H_0 = 73.04 \pm 1.04  $  km/s/Mpc  \cite{Riess:2021jrx}.
Hence, the $H_0$ tension has existed since the first
release of results from Planck Collaboration in 2013, which provided the 
estimation $H_0 = 67.40 \pm 1.40 
$ km/s/Mpc  \cite{Planck:2013pxb}. 
Meanwhile, the analysis of 
the nine-year Wilkinson 
Microwave Anisotropy Probe (WMAP) observations yielded a Hubble constant 
estimate of $H_0 = 70.00 \pm 2.20  $  km/s/Mpc  \cite{WMAP:2012fli}.  
In the following years, the 
improvement of the data and the reduction of errors, as well as the 
  consideration of additional datasets, has led to the worsening of the 
tension. For instance, as evidenced in 
\cite{Planck:2018vyg}, the Planck2018+lensing+BAO analysis results in $H_0 = 
67.66 \pm 0.42  $ km/s/Mpc. On the other hand,  an important   
  contribution came from recent observations from the James Webb Space 
Telescope, providing the strongest evidence yet that systematic errors in HST 
Cepheid measurements do not play a significant role in the present Hubble 
Tension \cite{Riess:2023bfx}. Finally, it is interesting to note the 
potentially intriguing situation,
where the American estimations of the current Hubble function often exceed
the 
European ones, which could be analyzed by future epistemologists.

If these tensions are not due to unknown 
systematics, they probably need a modification of standard lore in order to be 
alleviated. 
There are many directions one can follow in order achieve this.
 In particular,  
since
$
 \theta_s=\frac{r_s}{D_A},
$
where 
$
r_s
\propto \int_0^{t_{recom}}dt \frac{c_s(t)}{\rho (t)}
$
 is the sound horizon and 
$D_A\propto \frac{1}{H_0} \int_{t_{recom}}^{t_{today}}dt \frac{1}{\rho (t)}$
is the angular diameter 
distance, one could try to change either   $r_s$ or  $D_A$ or both. Solutions 
that affect $r_s$ are  referred to as ``early-time'' solutions, and 
solutions that alter  $D_A$ are   called ``late-time'' solutions.
Hence, in the literature one can find a large class of solutions, 
including modified gravity, 
early dark energy, extra relativistic degrees of freedom, bulk viscous models,
clustering dark energy, holographic dark energy, interacting dark energy, 
running vacuum models, Horndeski theories, decaying dark matter, 
string-inspired models, etc 
\cite{DiValentino:2021izs,DiValentino:2020zio,DiValentino:2015ola,  
Hu:2015rva,
Bernal:2016gxb,Kumar:2016zpg, Khosravi:2017hfi, 
DiValentino:2017iww,  DiValentino:2017oaw, 
DiValentino:2017zyq,
Sola:2017znb,Yang:2018euj,
DEramo:2018vss,Poulin:2018cxd,El-Zant:2018bsc,
Basilakos:2018arq,Adil:2021zxp,Nunes:2018xbm,Yang:2018prh,Pan:2019gop, 
Pan:2019jqh,Yan:2019gbw,DAgostino:2020dhv,
Anagnostopoulos:2020lec,Pandey:2019plg,Adhikari:2019fvb,
Benisty:2019pxb,Vagnozzi:2019ezj,Anagnostopoulos:2019miu,Braglia:2020auw,
Pan:2020bur, Capozziello:2020nyq,Saridakis:2019qwt, 
Escamilla-Rivera:2019ulu, DiValentino:2019jae,
Benevento:2020fev,Banerjee:2020xcn,Elizalde:2020mfs, 
DeFelice:2020cpt, Haridasu:2020pms,
Seto:2021xua,
Adi:2020qqf,Ballardini:2020iws,
LinaresCedeno:2020uxx,Odintsov:2020qzd,
Alestas:2021xes,Elizalde:2021kmo,Basilakos:2023xof,Krishnan:2021dyb,
Papanikolaou:2022did,
Theodoropoulos:2021hkk,Papanikolaou:2023crz} (for 
a 
review see \cite{Abdalla:2022yfr}).

In the present work we are interested in presenting a novel alleviation of both 
$H_0$ and $\sigma_8$ tension, obtained in the framework of Tsallis cosmology 
\cite{Lymperis:2018iuz}. Such a modified cosmological scenario arises from the 
application of the standard gravity-thermodynamics conjecture, namely the 
procedure to obtain the Friedmann equations from the first law of 
thermodynamics applied in the universe horizon 
\cite{Jacobson:1995ab,Padmanabhan:2003gd,Padmanabhan:2009vy}, but using Tsallis 
 non-additive entropy \cite{Tsallis:1987eu,Lyra:1997ggy,Wilk:1999dr}, instead 
of the usual Bekenstein-Hawking one.  Tsallis cosmology has been shown to lead 
to interesting cosmological phenomenology 
\cite{Sheykhi:2018dpn,Lalus:2019lzq,Nojiri:2019skr,Geng:2019shx,Ghoshal:2021ief,
Luciano:2021ndh,Zamora:2022cqz,Luciano:2022ely,Nojiri:2022dkr,Jizba:2022bfz}. 
Nevertheless, in the following we will show that 
 in such a framework we can obtain   phantom behavior for the effective 
dark-energy sector, which is one of the sufficient mechanisms that can 
alleviate the $H_0$ tension, as well as an increased friction term in the 
matter-perturbation evolution equation, which leads to 
smaller $\sigma_8$.

The plan of the work is the following: In Section \ref{themodel} we briefly 
review Tsallis cosmology, both at the background and perturbative levels. Then 
in Section \ref{Alleviatingtension} we show how Tsallis cosmology can lead to 
the alleviation of both $H_0$ and $\sigma_8$ tensions simultaneously.
Finally, in Section \ref{Conclusions} we summarize the obtained results.

\section{Modified cosmology through Tsallis entropy}
\label{themodel}

In this section we briefly review Tsallis cosmology. 
Tsallis non-additive entropy 
\cite{Tsallis:1987eu,Lyra:1997ggy,Wilk:1999dr}
 generalizes the standard thermodynamics to non-extensive one, and it possess 
the standard Boltzmann-Gibbs statistics as a limit. In a cosmological setup
this is quantified by an exponent $\delta$, and hence
the Tsallis entropy can be written in the form 
\cite{Tsallis:2012js}

\begin{equation}
\label{tsallisentr}
S_T=\frac{\tilde{\alpha}}{4G} A^{\delta}, 
\end{equation}
in units where $\hbar=k_B = c = 1$. In the above expression $G$ is the 
gravitational constant,
  $A\propto L^2$ is the area of the system with characteristic length $L$, 
 $\tilde{\alpha}$ is a positive constant with dimensions 
$[L^{2(1-\delta)}]$ and $\delta$ 
is the non-additivity parameter. As mentioned in the Introduction,
  in the 
case $\delta=1$ and $\tilde{\alpha}=1$, Tsallis entropy recovers the standard  
Bekenstein-Hawking additive entropy.

 We consider a Friedmann-Robertson-Walker (FRW)  metric of the form
\begin{equation}
ds^2=-dt^2+a^2(t)\left(\frac{dr^2}{1-kr^2}+r^2d\Omega^2 \right),
\label{metric}
\end{equation}
  with $a(t)$   the scale factor, and   $k=0,+1,-1$  the spatial curvature. 
We  substitute Tsallis entropy (\ref{tsallisentr}) into the first law of 
thermodynamics $-dE=TdS$, and we perform all the  steps of 
gravity-thermodynamics conjecture 
\cite{Jacobson:1995ab,Padmanabhan:2003gd,Padmanabhan:2009vy}. Specifically, we 
consider the boundary of the system to be  the Universe apparent horizon 
$\tilde{r}_a=(H^2+\frac{k}{a^2})^{-1}$, having temperature 
$T=1/(2\pi r_h)$, and being filled by the universe matter fluid, with energy 
density $\rho_m$ and pressure $p_m$ 
\cite{Cai:2005ra,Akbar:2006kj,Izquierdo:2005ku,
Cai:2006rs,Akbar:2006er,Paranjape:2006ca,Sheykhi:2007zp,
Jamil:2009eb,Cai:2009ph,
Wang:2009zv,
 Gim:2014nba, 
Fan:2014ala,Saridakis:2020lrg,Hernandez-Almada:2021rjs}. 
Hence, this
 leads to  (we consider only the more physically 
interesting case $\delta\neq2$)
\cite{Lymperis:2018iuz}
\be \label{tsgfe1}
-\frac{(4\pi)^{2-\delta}G}{\tilde{\alpha}}(\rho_m+p_m)=\delta 
\frac{\dot{H}-\frac{k}{a^2}}{\left(H^2+\frac{k}{
a^2}\right)^{\delta -1}},
\ee
and this by integration to
\be \label{tsgfe2}
\frac{2(4\pi)^{2-\delta}G}{3\tilde{\alpha}} \rho_m=\frac{ \delta 
}{2-\delta} \left(H^2+\frac{k}{a^2}\right) 
^{2-\delta}-\frac{\tilde{\Lambda}}{3\tilde{\alpha}},
\ee
with dots denoting time-derivatives, and
where $\tilde{\Lambda}$ is an integration constant, which can be considered as 
the cosmological constant.
Equations (\ref{tsgfe1}) and (\ref{tsgfe2}) are the 
two modified Friedmann equations for the non-extensive scenario with Tsallis 
entropy. Focusing on flat geometry, i.e. $k=0$, we can re-write them into the 
standard form 
\begin{eqnarray}
\label{FR1}
&&H^2=\frac{8\pi G}{3}\left(\rho_m+\rho_{DE}\right)\\
&&\dot{H}=-4\pi G \left(\rho_m+p_m+\rho_{DE}+p_{DE}\right),
\label{FR2}
\end{eqnarray}
where we have defined an effective dark energy density and pressure as 
\cite{Lymperis:2018iuz}
 \begin{eqnarray}
&&
\!\!\!\!\!\!\!\!\!\!\!\!\!\!\!\!\!\!
\rho_{DE}=\frac{3}{8\pi G} 
\left\{ \frac{\Lambda}{3}+H^2\left[1-\alpha \frac{ \delta}{ 2-\delta} 
H^{2(1-\delta) }
\right]
\right\},
\label{rhoDE1}
\end{eqnarray}
\begin{eqnarray}
&& \!\!\!\!\!\!\!\!\!\!\!\!\!\!\!\!\!\!\!\!
p_{DE}= -\frac{1}{8\pi G}\left\{
\Lambda
+2\dot{H}\left[1-\alpha\delta H^{2(1-\delta)}
\right] 
+3H^2\left[1-\alpha\frac{\delta}{2-\delta}H^{
2(1-\delta)}
\right]
\right\},
\label{pDE1}
\end{eqnarray}
as well as the new constants 
$ \Lambda  \equiv (4\pi)^{\delta-1}\tilde{\Lambda}$ and 
$\alpha\equiv (4\pi)^{\delta-1}\tilde{\alpha}$. In these lines, the 
 equation-of-state parameter for the effective dark 
energy
is
\begin{eqnarray}
w_{DE}\equiv\frac{p_{DE}}{\rho_{DE}}=-1-
\frac{     
  2\dot{H}\left[1-\alpha\delta H^{2(1-\delta)}
\right]
 }{\Lambda+3H^2\left[1-\frac{\alpha\delta}{2-\delta}H^{2(1-\delta)}
\right]}
\label{wDE}.
\end{eqnarray}
We mention that  for $\delta =1$ and $ \alpha =1$ the above 
expressions recover the standard ones as expected.

Let us elaborate the aforementioned equations. For simplicity we consider dust 
matter ($p_m=0$) and we introduce the  
density parameters 
through
 \begin{eqnarray} \label{omatter}
&&\Omega_m=\frac{8\pi G}{3H^2} \rho_m\\
&& \label{ode}
\Omega_{DE}=\frac{8\pi G}{3H^2} \rho_{DE}.
 \end{eqnarray} 
Doing so, 
  the Hubble parameter can be written as
\be \label{h2}
H=\frac{\sqrt{\Omega_{m0}} H_{0}}{\sqrt{a^3 (1-\Omega_{DE})}},
\ee
where ``0" denotes the present value of the corresponding quantity.  
In the following it proves more convenient to introduce  the 
redshift $z$, defined  as  $ 
 1+z=1/a$, having imposed the present scale factor to 1.
Substituting (\ref{rhoDE1}) into (\ref{ode}) and taking into account (\ref{h2}) 
we acquire 
 \begin{eqnarray} 
 \label{omegaDEtsal}
&&
\!\!\!\!\!\!\!\!\!\!\!\!\!\!
\Omega_{DE}(z)=
1-H^{2}_{0}\Omega_{m0}(1+z)^3 
 \left\{\frac{
(2\!-\!\delta)}{\alpha 
\delta}\left[H^{2}_{0}\Omega_{m0}(1\!+\!z)^3+\frac{\Lambda}{3}
\right]\right\}^{\frac{1}{\delta -2}},
 \end{eqnarray}
while from (\ref{wDE}) we find \cite{Lymperis:2018iuz}
\be
\label{wDEfinal}
w_{DE}(z)=-1+\frac{\left\{3[1-\Omega_{DE}(z)]+(1+z)\Omega_{DE}'(z)\right\}
\left\{1-\alpha 
\delta 
\left[\frac{
H^{2}_{0}\Omega_{m0}(1+z)^3}{1-\Omega_{DE}(z)}\right]^{1-\delta}\right\}}{[
1-\Omega_{DE}
(z)]\left\{\frac{\Lambda 
[1-\Omega_{DE}(z)]}{H^{2}_{0}\Omega_{m0}(1+z)^3}+3\left\{1-\frac{\alpha 
\delta}{2-\delta}\left[\frac{H^{2}_{0}\Omega_{m0}(1+z)^3}{1-\Omega_{DE}(z)}
\right]^{ 
1-\delta}\right\} \right\}},
\ee
where primes denote differentiation with respect to $z$. 
Finally, note that applying   (\ref{omegaDEtsal}) at present time ($z=0$)
we acquire a relation of the parameters, namely
\begin{eqnarray} \label{lambda}
 \Lambda=\frac{3\alpha\delta}{2-\delta}H_0^{2(2-\delta)}-3H_0^2\Omega_{m0},
\end{eqnarray}
from which we deduce that our model has  two extra free parameters, namely 
$\alpha$ and $\delta$.

We close this section by presenting the behavior of Tsallis cosmology at the 
perturbative level. Introducing as usual the matter overdensity 
$\delta_m:=\delta\rho_m/\rho_m$, and focusing without loss of generality on the 
case $\alpha =1$, one can show that  its evolution equation 
is given by  \cite{Sheykhi:2022gzb}
 \begin{equation}
 \label{deltaz}
\delta^{\prime\prime}_{m}
+\frac{2\left (4\!-\!2\delta \right)-
\left(9\!-\!6\delta\! +\!8\pi G\Lambda H^{2\delta\! -\!4} \right )}{\left 
(4\!-\!2\delta \right)(1+z)}\delta^{\prime}_{m}
+
\frac{3^{\frac{1}{\delta -2}}\!
 \left [\left (1\!-\!2\delta 
\right)\rho^{\frac{1}{2-\delta}}_{m}-9\left(1\!-\!\delta \right)\Lambda  
\rho^{\frac{\delta -1}{2\!-\!\delta}}_{m}\right ]8\pi G 
}{2\left(2-\delta \right)^{2}H^{2}(1+z)^{2}}\delta_{m}=0,
\end{equation}
 with $\Lambda$ given by (\ref{lambda}). Note that comparing to standard 
result,  in the above expression we have a different friction term (the 
second term), as well as an effective  Newton's constant (the last term). 
Clearly, in the case $\delta=1$, one 
recovers the standard result, which in the case of matter domination 
($\Omega_m\approx1$) gives
\begin{equation}
 \delta^{\prime\prime}_{m}+\frac{1}{2(1+z)}\delta^{\prime}_{m}
-\frac{3}{2(1+z)^2 } \delta_{m}=0
\end{equation}
as expected. Lastly, after
obtaining the solution for $\delta_m(z)$, we can calculate the physically 
interesting observational quantity 
 \begin{equation}
f\sigma_8(z)=f(z) \,\sigma(z),
 \end{equation}
 with 
$f(z):=-\frac{d\ln\delta_m(z)}{d\ln z}$ and 
$\sigma(z):=\sigma_8\frac{\delta_m(z)}{\delta_m(0)}$ \cite{Abdalla:2022yfr}.

\section{Alleviating $H_{0}$ and $\sigma_{8}$ tensions }
\label{Alleviatingtension}

Let us now investigate how the scenario of Tsallis cosmology can alleviate  both
$H_{0}$ and $\sigma_{8}$ tensions. As we observe, the Tsallis exponent $\delta$ 
affects the background evolution, as well as the perturbation behavior.
As it was in \cite{Abdalla:2022yfr} and in 
\cite{Heisenberg:2022lob,Heisenberg:2022gqk}, one of the  efficient mechanisms 
that can alleviate the $H_0$ tension is to obtain an effective dark-energy 
equation-of-state parameter lying in the phantom regime, while one of the 
efficient mechanisms that can alleviate the $\sigma_{8}$  tension is to obtain 
an increased friction term or a smaller effective Newton's constant in the 
evolution equation of $\delta_m$.
Hence, our strategy will be to choose $\delta$ in order to fulfill the above 
requirements.  

We start from the fact that  in $\Lambda$CDM cosmology the Hubble function is 
providing by the relation
\be
\label{hlcdm}
H_{\Lambda CDM}(z)\equiv H_{0}\sqrt{\Omega_{m0}(1+z)^{3}+1-\Omega_{m0}}.
\ee
On the other hand, in Tsallis cosmology the Hubble function is given by 
(\ref{h2}) with $\Omega_{DE}(z)$ provided by (\ref{omegaDEtsal}). Hence, we can 
choose model parameters which our $H(z)$  coincides with $H_{\Lambda 
\text{CDM}}(z)$ of  (\ref{hlcdm}) at $z= z_{\rm CMB}\approx 1100$,
namely $H(z\rightarrow z_{\rm CMB}) \approx H_{\Lambda\text{CDM}}(z\rightarrow 
z_{\rm CMB})$,
but give 
$H(z\rightarrow 0) > H_{\Lambda\text{CDM}}(z\rightarrow 0)$. Finally, as usual, 
we desire to have the standard evolution of $\Omega_m$ and $\Omega_{DE}$, with 
the sequence of matter and dark-energy epochs, and with 
  $\Omega_{m0} \approx 
0.31$, according to observations \cite{Planck:2018vyg}. We mention here that 
the range of values $\{\delta,\Omega_{m0}\}$ that we are using is well within 
the observational bounds in these kinds of theories 
\cite{Lymperis:2018iuz,Asghari:2021lzu}.

\begin{figure}[ht]
\centering
\includegraphics[width=9cm]{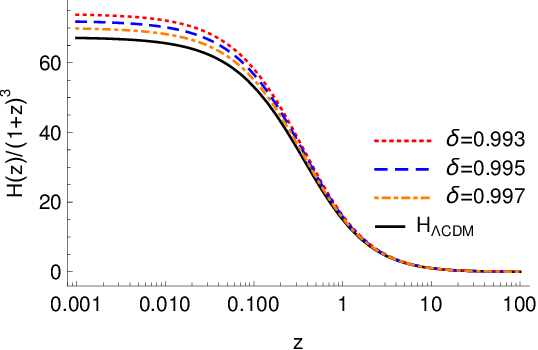}
\caption{{\it The normalized $H(z)/(1 + z)^{3}$ in units of
$km/s/Mpc$ as a function of the redshift, for $\Lambda$CDM cosmology 
(black-solid) and for Tsallis cosmology for $\alpha =1$  and for  
$\delta=0.993$  (red 
- dotted),  $\delta=0.995$ (blue - dashed) and $\delta=0.997$ (orange - 
dashed-dotted). We have imposed $\Omega_{m0} \approx 
0.31$.}}
\label{fig:fig2}
\end{figure}

In Fig. \ref{fig:fig2} we depict the normalized $H(z)/(1+z)^{3}$ as a function 
of the redshift parameter, for $\Lambda$CDM scenario, as well as    for 
 Tsallis cosmology for various values of the entropic exponent $\delta$. As we 
observe, for small deviations of $\delta$ below the standard entropy value   
$\delta=1$ we can have a coincidence to $\Lambda$CDM cosmology at 
 high and intermediate redshifts, while at small redshifts the modified Tsallis 
scenario stabilizes in higher values of $H_{0}$. More specifically, the value 
of 
$H_{0}$ depends on  $\delta$, and 
it can be around $H_{0} \approx 74$ km/s/Mpc for $\delta = 0.993$. Hence, 
Tsallis cosmology, with $\delta$ values slightly less than the standard value, 
can indeed alleviate $H_0$ tension ($\delta>1$ values do not lead to 
alleviation). 
\begin{figure}[!]
\centering
\includegraphics[width=9cm]{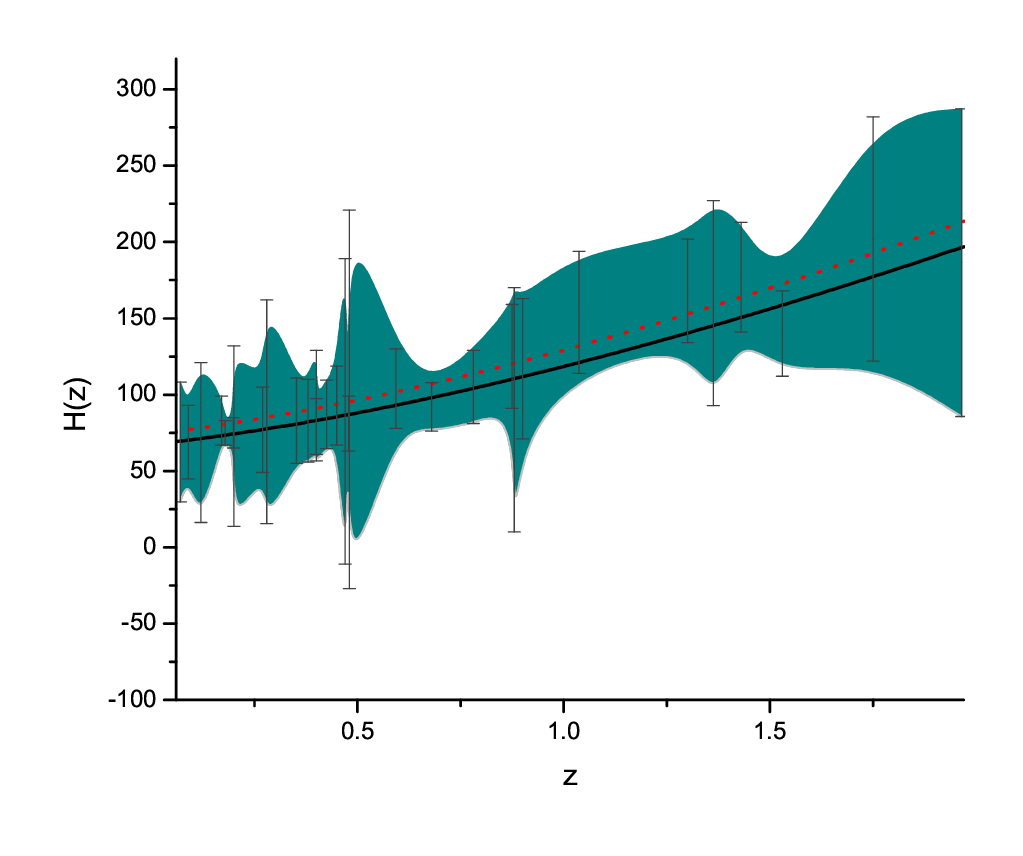}
\caption{{\it The $H(z)$ evolution in units of km/s/Mpc as a function of
the redshift, for $\Lambda$CDM scenario (black line) and for Tsallis cosmology 
with $\alpha =1$  and 
 $\delta =0.993$ (red dotted line), on top of the CC data points   
 at 2$\sigma$ confidence level \cite{Yu:2017iju}.
We have imposed $\Omega_{m0} \approx 0.31$.}}
\label{fig:fig3}
\end{figure} 

As an additional verification, we confront the obtained evolution with 
 Cosmic Chronometer (CC) 
data   \cite{Jimenez:2001gg}, namely    datasets
 based on $H(z)$    measurements through the relative ages of massive 
passively evolving galaxies  \cite{Yu:2017iju}. In Fig. 
\ref{fig:fig3} we present $H(z)$ for $\Lambda$CDM scenario and for Tsallis 
cosmology, on top of  CC
data from \cite{Yu:2017iju} at 2$\sigma$ confidence level. As we see, the 
agreement is very good, $ H(z)$  of Tsallis cosmology lies within 
the    CC data, exhibiting 
a slightly higher accelerating behavior at low redshifts, for
the parameter sets $\{\Omega_{m0}, \delta\} = \{0.31, 0.993\}$.
\begin{figure}[!h]
\centering
\includegraphics[width=8cm]{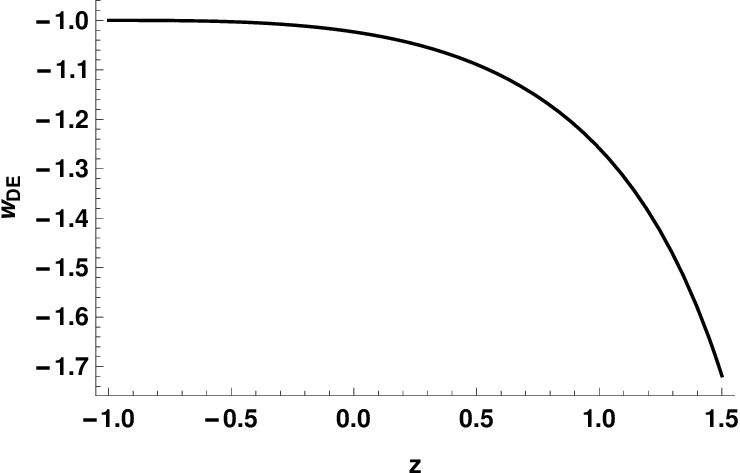}
\caption{{\it{ The evolution of the dark-energy
equation-of-state parameter $w_{DE}$  as a function of
the redshift,    for Tsallis cosmology  with $\alpha =1$ and $\delta 
=0.993$. } }}
\label{fig:fig1}
\end{figure} 

 \begin{figure}[!h]
\centering
\includegraphics[width=10cm]{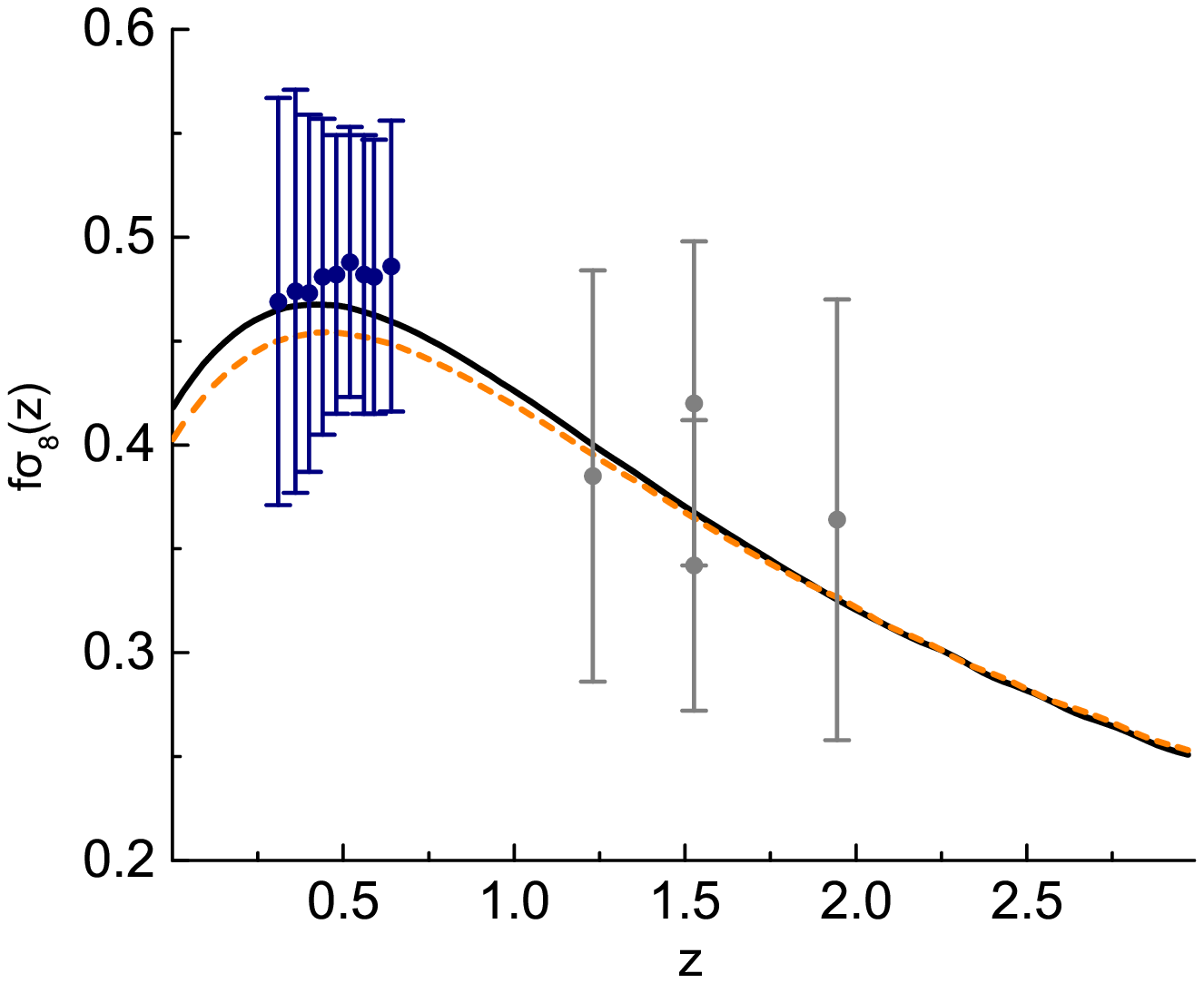}
\caption{{\it Evolution of f$\sigma_8$ in $\Lambda$CDM scenario (black solid) 
and in Tsallis cosmology with $\alpha =1$  and 
 $\delta =0.993$ (orange dashed). The blue data points are from Baryonic 
Acoustic Oscillations 
(BAO) observations in SDSS-III DR12 \cite{Wang:2017wia}, while the gray data 
points at higher redshifts are from SDSS-IV DR14 
\cite{Gil-Marin:2018cgo,Hou:2018yny,Zhao:2018gvb}.}}
\label{fig:fig4}
\end{figure} 

Let us now examine what is the mechanism behind the $H_0$ tension alleviation.
In Fig. \ref{fig:fig1} we present the evolution of the dark-energy 
equation-of-state parameter (\ref{wDEfinal}), for the value of the parameter  
$\delta$   that 
alleviates the tension. As we observe, it lies in the phantom regime, which as 
we mentioned above is one of the sufficient ways to alleviate the tension 
\cite{Heisenberg:2022lob,Heisenberg:2022gqk}. Note 
that according to (\ref{wDEfinal}) in principle $w_{DE}(z)$ can be 
quintessence-like,  phantom-like, or experience the phantom-divide crossing 
during the evolution, according to the value of $\delta$, and this is the 
reason we restricted ourselves to values $\delta\lesssim1$, since these are 
needed to obtain the correct amount of phantom behavior (concerning other 
potential astrophysical implications of phantom behavior see 
\cite{Cai:2009zp}).
Finally, note that we choose to extend    Fig. \ref{fig:fig1}  to 
negative redshifts, namely to the future, in order to show 
 that in the scenario at hand   as time passes the phantom 
behavior becomes less and less significant, and that the dark energy 
sector will asymptotically behave as a cosmological constant.

We proceed to the examination of the $\sigma_8$ tension. As we mentioned above, 
the evolution equation of matter overdensity $\delta_m$ is given by 
(\ref{deltaz}). 
In Fig. \ref{fig:fig4}, we present the
evolution of $f\sigma_{8}$ for $\Lambda$CDM scenario, as well as for   Tsallis 
cosmology, on top of observational data. As we observe, Tsallis cosmology can 
indeed reduce  $f\sigma_{8}$ and alleviate $\sigma_8$ tension too, for the 
same parameter choice that can alleviate $H_0$ tension. This simultaneous 
alleviation of the tensions is not easy to be obtained in alternative 
cosmological scenarios, and it is the main result of the present work.

Finally, let us examine the mechanism behind the $\sigma_8$ tension 
alleviation. As we observe from (\ref{deltaz}), the scenario at hand has  a 
different friction term   as well as an effective  Newton's constant. One can 
see that under the above parameter choice, we obtain an increased friction and 
an effective Newton's constant smaller than the usual one. And this is indeed 
one of the sufficient mechanisms to alleviate  $\sigma_8$ tension 
\cite{Heisenberg:2022lob,Heisenberg:2022gqk}.

\section{Conclusions}
\label{Conclusions}

We presented how Tsallis cosmology can alleviate both   $H_0$ and $\sigma_8$  
tensions simultaneously. Such a modified cosmological scenario is obtained by 
the application of the gravity-thermodynamics conjecture, but using the 
non-additive Tsallis entropy, instead of the standard Bekenstein-Hawking one. 
Hence, one obtains modified Friedmann equations, with extra terms that depend 
on the new Tsallis exponent $\delta$ that quantifies the departure from 
standard entropy.

In Tsallis cosmology one acquires an effective dark-energy sector with 
equation-of-state parameter that can be quintessence-like or phantom-like. 
Additionally, at the perturbative level one extracts the evolution equation for 
matter overdensity, with extra terms in the friction term as well as in the 
effective Newton's constant.

As we showed, for particular choice of the Tsallis parameter $\delta$ we can 
obtain a phantom effective dark energy, which is known to be one of the 
sufficient mechanisms that can alleviate  $H_0$ tension. Interestingly enough, 
for the same parameter choice we obtain  an increased friction term and 
an effective Newton's constant smaller than the usual one, and thus the 
$\sigma_8$  
tension is also solved.

In summary, Tsallis cosmology can simultaneously alleviate both   $H_0$ and 
$\sigma_8$   tensions. Note that in general this is not easily obtained in 
alternative cosmological scenarios \cite{Abdalla:2022yfr}. Even within the 
framework of other modified entropic cosmologies this cannot be acquired, such 
as in the case of Barrow cosmology \cite{Saridakis:2020lrg,Barrow:2020kug} or 
Kaniadakis cosmology \cite{Lymperis:2021qty}, due to the less freedom one has 
in choosing the values of the extra entropic parameters, namely $\Delta$ and 
$K$ respectively (e.g. in Barrow cosmology $H_0$ alleviation would 
mathematically need $\Delta\lesssim0$ which is not allowed). Thus, this feature 
acts as a significant advantage of Tsallis modified cosmology.

\acknowledgments
 M.P. is supported by the Basic Research program
of the National Technical University of Athens (NTUA,
PEVE) 65232600-ACT-MTG: Alleviating Cosmological
Tensions Through Modified Theories of Gravity. The authors acknowledge the 
contribution of the LISA CosWG, and of   COST Actions  CA18108 
``Quantum Gravity 
Phenomenology in the multi-messenger approach''  and 
CA21136 ``Addressing observational tensions in cosmology with systematics and 
fundamental physics (CosmoVerse)'' A.L. would like to thank ``Private Maternity 
of Patras'' for the hospitality during the preparation of a part of the present 
manuscript, where his daughter was born.


\bibliographystyle{JHEP} 
\bibliography{paperRefs}

\providecommand{\href}[2]{#2}\begingroup\raggedright\begin{thebibliography}{100}

\bibitem{Perivolaropoulos:2021jda}
L.~Perivolaropoulos and F.~Skara, \emph{{Challenges for
  \ensuremath{\Lambda}CDM: An update}},
  \href{https://doi.org/10.1016/j.newar.2022.101659}{\emph{New Astron. Rev.}
  {\bfseries 95} (2022) 101659}
  [\href{https://arxiv.org/abs/2105.05208}{{\ttfamily 2105.05208}}].

\bibitem{Planck:2018vyg}
{\scshape Planck} collaboration, \emph{{Planck 2018 results. VI. Cosmological
  parameters}},
  \href{https://doi.org/10.1051/0004-6361/201833910}{\emph{Astron. Astrophys.}
  {\bfseries 641} (2020) A6}
  [\href{https://arxiv.org/abs/1807.06209}{{\ttfamily 1807.06209}}].

\bibitem{Zarrouk:2018vwy}
P.~Zarrouk et~al., \emph{{The clustering of the SDSS-IV extended Baryon
  Oscillation Spectroscopic Survey DR14 quasar sample: measurement of the
  growth rate of structure from the anisotropic correlation function between
  redshift 0.8 and 2.2}},
  \href{https://doi.org/10.1093/mnras/sty506}{\emph{Mon. Not. Roy. Astron.
  Soc.} {\bfseries 477} (2018) 1639}
  [\href{https://arxiv.org/abs/1801.03062}{{\ttfamily 1801.03062}}].

\bibitem{BOSS:2016wmc}
{\scshape BOSS} collaboration, \emph{{The clustering of galaxies in the
  completed SDSS-III Baryon Oscillation Spectroscopic Survey: cosmological
  analysis of the DR12 galaxy sample}},
  \href{https://doi.org/10.1093/mnras/stx721}{\emph{Mon. Not. Roy. Astron.
  Soc.} {\bfseries 470} (2017) 2617}
  [\href{https://arxiv.org/abs/1607.03155}{{\ttfamily 1607.03155}}].

\bibitem{HST:2000azd}
{\scshape HST} collaboration, \emph{{Final results from the Hubble Space
  Telescope key project to measure the Hubble constant}},
  \href{https://doi.org/10.1086/320638}{\emph{Astrophys. J.} {\bfseries 553}
  (2001) 47} [\href{https://arxiv.org/abs/astro-ph/0012376}{{\ttfamily
  astro-ph/0012376}}].

\bibitem{Freedman:2012ny}
W.L.~Freedman, B.F.~Madore, V.~Scowcroft, C.~Burns, A.~Monson, S.E.~Persson
  et~al., \emph{{Carnegie Hubble Program: A Mid-Infrared Calibration of the
  Hubble Constant}},
  \href{https://doi.org/10.1088/0004-637X/758/1/24}{\emph{Astrophys. J.}
  {\bfseries 758} (2012) 24} [\href{https://arxiv.org/abs/1208.3281}{{\ttfamily
  1208.3281}}].

\bibitem{Riess:2021jrx}
A.G.~Riess et~al., \emph{{A Comprehensive Measurement of the Local Value of the
  Hubble Constant with Uncertainty from the Hubble Space Telescope and the
  SH0ES Team}},
  \href{https://doi.org/10.3847/2041-8213/ac5c5b}{\emph{Astrophys. J. Lett.}
  {\bfseries 934} (2022) L7}
  [\href{https://arxiv.org/abs/2112.04510}{{\ttfamily 2112.04510}}].

\bibitem{Planck:2013pxb}
{\scshape Planck} collaboration, \emph{{Planck 2013 results. XVI. Cosmological
  parameters}},
  \href{https://doi.org/10.1051/0004-6361/201321591}{\emph{Astron. Astrophys.}
  {\bfseries 571} (2014) A16}
  [\href{https://arxiv.org/abs/1303.5076}{{\ttfamily 1303.5076}}].

\bibitem{WMAP:2012fli}
{\scshape WMAP} collaboration, \emph{{Nine-Year Wilkinson Microwave Anisotropy
  Probe (WMAP) Observations: Final Maps and Results}},
  \href{https://doi.org/10.1088/0067-0049/208/2/20}{\emph{Astrophys. J. Suppl.}
  {\bfseries 208} (2013) 20} [\href{https://arxiv.org/abs/1212.5225}{{\ttfamily
  1212.5225}}].

\bibitem{Riess:2023bfx}
A.G.~Riess, G.S.~Anand, W.~Yuan, S.~Casertano, A.~Dolphin, L.M.~Macri et~al.,
  \emph{{Crowded No More: The Accuracy of the Hubble Constant Tested with
  High-resolution Observations of Cepheids by JWST}},
  \href{https://doi.org/10.3847/2041-8213/acf769}{\emph{Astrophys. J. Lett.}
  {\bfseries 956} (2023) L18}
  [\href{https://arxiv.org/abs/2307.15806}{{\ttfamily 2307.15806}}].

\bibitem{DiValentino:2021izs}
E.~Di~Valentino, O.~Mena, S.~Pan, L.~Visinelli, W.~Yang, A.~Melchiorri et~al.,
  \emph{{In the realm of the Hubble tension\textemdash{}a review of
  solutions}}, \href{https://doi.org/10.1088/1361-6382/ac086d}{\emph{Class.
  Quant. Grav.} {\bfseries 38} (2021) 153001}
  [\href{https://arxiv.org/abs/2103.01183}{{\ttfamily 2103.01183}}].

\bibitem{DiValentino:2020zio}
E.~Di~Valentino et~al., \emph{{Snowmass2021 - Letter of interest cosmology
  intertwined II: The hubble constant tension}},
  \href{https://doi.org/10.1016/j.astropartphys.2021.102605}{\emph{Astropart.
  Phys.} {\bfseries 131} (2021) 102605}
  [\href{https://arxiv.org/abs/2008.11284}{{\ttfamily 2008.11284}}].

\bibitem{DiValentino:2015ola}
E.~Di~Valentino, A.~Melchiorri and J.~Silk, \emph{{Beyond six parameters:
  extending $\Lambda$CDM}},
  \href{https://doi.org/10.1103/PhysRevD.92.121302}{\emph{Phys. Rev. D}
  {\bfseries 92} (2015) 121302}
  [\href{https://arxiv.org/abs/1507.06646}{{\ttfamily 1507.06646}}].

\bibitem{Hu:2015rva}
B.~Hu and M.~Raveri, \emph{{Can modified gravity models reconcile the tension
  between the CMB anisotropy and lensing maps in Planck-like observations?}},
  \href{https://doi.org/10.1103/PhysRevD.91.123515}{\emph{Phys. Rev. D}
  {\bfseries 91} (2015) 123515}
  [\href{https://arxiv.org/abs/1502.06599}{{\ttfamily 1502.06599}}].

\bibitem{Bernal:2016gxb}
J.L.~Bernal, L.~Verde and A.G.~Riess, \emph{{The trouble with $H_0$}},
  \href{https://doi.org/10.1088/1475-7516/2016/10/019}{\emph{JCAP} {\bfseries
  10} (2016) 019} [\href{https://arxiv.org/abs/1607.05617}{{\ttfamily
  1607.05617}}].

\bibitem{Kumar:2016zpg}
S.~Kumar and R.C.~Nunes, \emph{{Probing the interaction between dark matter and
  dark energy in the presence of massive neutrinos}},
  \href{https://doi.org/10.1103/PhysRevD.94.123511}{\emph{Phys. Rev. D}
  {\bfseries 94} (2016) 123511}
  [\href{https://arxiv.org/abs/1608.02454}{{\ttfamily 1608.02454}}].

\bibitem{Khosravi:2017hfi}
N.~Khosravi, S.~Baghram, N.~Afshordi and N.~Altamirano, \emph{{$H_0$ tension as
  a hint for a transition in gravitational theory}},
  \href{https://doi.org/10.1103/PhysRevD.99.103526}{\emph{Phys. Rev. D}
  {\bfseries 99} (2019) 103526}
  [\href{https://arxiv.org/abs/1710.09366}{{\ttfamily 1710.09366}}].

\bibitem{DiValentino:2017iww}
E.~Di~Valentino, A.~Melchiorri and O.~Mena, \emph{{Can interacting dark energy
  solve the $H_0$ tension?}},
  \href{https://doi.org/10.1103/PhysRevD.96.043503}{\emph{Phys. Rev. D}
  {\bfseries 96} (2017) 043503}
  [\href{https://arxiv.org/abs/1704.08342}{{\ttfamily 1704.08342}}].

\bibitem{DiValentino:2017oaw}
E.~Di~Valentino, C.~B\o{}ehm, E.~Hivon and F.R.~Bouchet, \emph{{Reducing the
  $H_0$ and $\sigma_8$ tensions with Dark Matter-neutrino interactions}},
  \href{https://doi.org/10.1103/PhysRevD.97.043513}{\emph{Phys. Rev. D}
  {\bfseries 97} (2018) 043513}
  [\href{https://arxiv.org/abs/1710.02559}{{\ttfamily 1710.02559}}].

\bibitem{DiValentino:2017zyq}
E.~Di~Valentino, A.~Melchiorri, E.V.~Linder and J.~Silk, \emph{{Constraining
  Dark Energy Dynamics in Extended Parameter Space}},
  \href{https://doi.org/10.1103/PhysRevD.96.023523}{\emph{Phys. Rev. D}
  {\bfseries 96} (2017) 023523}
  [\href{https://arxiv.org/abs/1704.00762}{{\ttfamily 1704.00762}}].

\bibitem{Sola:2017znb}
J.~Sol\`a, A.~G\'omez-Valent and J.~de~Cruz~P\'erez, \emph{{The $H_0$ tension
  in light of vacuum dynamics in the Universe}},
  \href{https://doi.org/10.1016/j.physletb.2017.09.073}{\emph{Phys. Lett. B}
  {\bfseries 774} (2017) 317}
  [\href{https://arxiv.org/abs/1705.06723}{{\ttfamily 1705.06723}}].

\bibitem{Yang:2018euj}
W.~Yang, S.~Pan, E.~Di~Valentino, R.C.~Nunes, S.~Vagnozzi and D.F.~Mota,
  \emph{{Tale of stable interacting dark energy, observational signatures, and
  the $H_0$ tension}},
  \href{https://doi.org/10.1088/1475-7516/2018/09/019}{\emph{JCAP} {\bfseries
  09} (2018) 019} [\href{https://arxiv.org/abs/1805.08252}{{\ttfamily
  1805.08252}}].

\bibitem{DEramo:2018vss}
F.~D'Eramo, R.Z.~Ferreira, A.~Notari and J.L.~Bernal, \emph{{Hot Axions and the
  $H_0$ tension}},
  \href{https://doi.org/10.1088/1475-7516/2018/11/014}{\emph{JCAP} {\bfseries
  11} (2018) 014} [\href{https://arxiv.org/abs/1808.07430}{{\ttfamily
  1808.07430}}].

\bibitem{Poulin:2018cxd}
V.~Poulin, T.L.~Smith, T.~Karwal and M.~Kamionkowski, \emph{{Early Dark Energy
  Can Resolve The Hubble Tension}},
  \href{https://doi.org/10.1103/PhysRevLett.122.221301}{\emph{Phys. Rev. Lett.}
  {\bfseries 122} (2019) 221301}
  [\href{https://arxiv.org/abs/1811.04083}{{\ttfamily 1811.04083}}].

\bibitem{El-Zant:2018bsc}
A.~El-Zant, W.~El~Hanafy and S.~Elgammal, \emph{{$H_0$ Tension and the Phantom
  Regime: A Case Study in Terms of an Infrared $f(T)$ Gravity}},
  \href{https://doi.org/10.3847/1538-4357/aafa12}{\emph{Astrophys. J.}
  {\bfseries 871} (2019) 210}
  [\href{https://arxiv.org/abs/1809.09390}{{\ttfamily 1809.09390}}].

\bibitem{Basilakos:2018arq}
S.~Basilakos, S.~Nesseris, F.K.~Anagnostopoulos and E.N.~Saridakis,
  \emph{{Updated constraints on $f(T)$ models using direct and indirect
  measurements of the Hubble parameter}},
  \href{https://doi.org/10.1088/1475-7516/2018/08/008}{\emph{JCAP} {\bfseries
  08} (2018) 008} [\href{https://arxiv.org/abs/1803.09278}{{\ttfamily
  1803.09278}}].

\bibitem{Adil:2021zxp}
S.A.~Adil, M.R.~Gangopadhyay, M.~Sami and M.K.~Sharma, \emph{{Late-time
  acceleration due to a generic modification of gravity and the Hubble
  tension}}, \href{https://doi.org/10.1103/PhysRevD.104.103534}{\emph{Phys.
  Rev. D} {\bfseries 104} (2021) 103534}
  [\href{https://arxiv.org/abs/2106.03093}{{\ttfamily 2106.03093}}].

\bibitem{Nunes:2018xbm}
R.C.~Nunes, \emph{{Structure formation in $f(T)$ gravity and a solution for
  $H_0$ tension}},
  \href{https://doi.org/10.1088/1475-7516/2018/05/052}{\emph{JCAP} {\bfseries
  05} (2018) 052} [\href{https://arxiv.org/abs/1802.02281}{{\ttfamily
  1802.02281}}].

\bibitem{Yang:2018prh}
W.~Yang, S.~Pan, E.~Di~Valentino and E.N.~Saridakis, \emph{{Observational
  constraints on dynamical dark energy with pivoting redshift}},
  \href{https://doi.org/10.3390/universe5110219}{\emph{Universe} {\bfseries 5}
  (2019) 219} [\href{https://arxiv.org/abs/1811.06932}{{\ttfamily
  1811.06932}}].

\bibitem{Pan:2019gop}
S.~Pan, W.~Yang, E.~Di~Valentino, E.N.~Saridakis and S.~Chakraborty,
  \emph{{Interacting scenarios with dynamical dark energy: Observational
  constraints and alleviation of the $H_0$ tension}},
  \href{https://doi.org/10.1103/PhysRevD.100.103520}{\emph{Phys. Rev. D}
  {\bfseries 100} (2019) 103520}
  [\href{https://arxiv.org/abs/1907.07540}{{\ttfamily 1907.07540}}].

\bibitem{Pan:2019jqh}
S.~Pan, W.~Yang, C.~Singha and E.N.~Saridakis, \emph{{Observational constraints
  on sign-changeable interaction models and alleviation of the $H_0$ tension}},
  \href{https://doi.org/10.1103/PhysRevD.100.083539}{\emph{Phys. Rev. D}
  {\bfseries 100} (2019) 083539}
  [\href{https://arxiv.org/abs/1903.10969}{{\ttfamily 1903.10969}}].

\bibitem{Yan:2019gbw}
S.-F.~Yan, P.~Zhang, J.-W.~Chen, X.-Z.~Zhang, Y.-F.~Cai and E.N.~Saridakis,
  \emph{{Interpreting cosmological tensions from the effective field theory of
  torsional gravity}},
  \href{https://doi.org/10.1103/PhysRevD.101.121301}{\emph{Phys. Rev. D}
  {\bfseries 101} (2020) 121301}
  [\href{https://arxiv.org/abs/1909.06388}{{\ttfamily 1909.06388}}].

\bibitem{DAgostino:2020dhv}
R.~D'Agostino and R.C.~Nunes, \emph{{Measurements of $H_0$ in modified gravity
  theories: The role of lensed quasars in the late-time Universe}},
  \href{https://doi.org/10.1103/PhysRevD.101.103505}{\emph{Phys. Rev. D}
  {\bfseries 101} (2020) 103505}
  [\href{https://arxiv.org/abs/2002.06381}{{\ttfamily 2002.06381}}].

\bibitem{Anagnostopoulos:2020lec}
F.K.~Anagnostopoulos, S.~Basilakos and E.N.~Saridakis, \emph{{Observational
  constraints on Myrzakulov gravity}},
  \href{https://doi.org/10.1103/PhysRevD.103.104013}{\emph{Phys. Rev. D}
  {\bfseries 103} (2021) 104013}
  [\href{https://arxiv.org/abs/2012.06524}{{\ttfamily 2012.06524}}].

\bibitem{Pandey:2019plg}
K.L.~Pandey, T.~Karwal and S.~Das, \emph{{Alleviating the $H_0$ and $\sigma_8$
  anomalies with a decaying dark matter model}},
  \href{https://doi.org/10.1088/1475-7516/2020/07/026}{\emph{JCAP} {\bfseries
  07} (2020) 026} [\href{https://arxiv.org/abs/1902.10636}{{\ttfamily
  1902.10636}}].

\bibitem{Adhikari:2019fvb}
S.~Adhikari and D.~Huterer, \emph{{Super-CMB fluctuations and the Hubble
  tension}}, \href{https://doi.org/10.1016/j.dark.2020.100539}{\emph{Phys. Dark
  Univ.} {\bfseries 28} (2020) 100539}
  [\href{https://arxiv.org/abs/1905.02278}{{\ttfamily 1905.02278}}].

\bibitem{Benisty:2019pxb}
D.~Benisty, \emph{{Cosmology of fermionic dark energy coupled to curvature}},
  \href{https://doi.org/10.1016/j.nuclphysb.2023.116251}{\emph{Nucl. Phys. B}
  {\bfseries 992} (2023) 116251}
  [\href{https://arxiv.org/abs/1912.11124}{{\ttfamily 1912.11124}}].

\bibitem{Vagnozzi:2019ezj}
S.~Vagnozzi, \emph{{New physics in light of the $H_0$ tension: An alternative
  view}}, \href{https://doi.org/10.1103/PhysRevD.102.023518}{\emph{Phys. Rev.
  D} {\bfseries 102} (2020) 023518}
  [\href{https://arxiv.org/abs/1907.07569}{{\ttfamily 1907.07569}}].

\bibitem{Anagnostopoulos:2019miu}
F.K.~Anagnostopoulos, S.~Basilakos and E.N.~Saridakis, \emph{{Bayesian analysis
  of $f(T)$ gravity using $f\sigma_8$ data}},
  \href{https://doi.org/10.1103/PhysRevD.100.083517}{\emph{Phys. Rev. D}
  {\bfseries 100} (2019) 083517}
  [\href{https://arxiv.org/abs/1907.07533}{{\ttfamily 1907.07533}}].

\bibitem{Braglia:2020auw}
M.~Braglia, M.~Ballardini, F.~Finelli and K.~Koyama, \emph{{Early modified
  gravity in light of the $H_0$ tension and LSS data}},
  \href{https://doi.org/10.1103/PhysRevD.103.043528}{\emph{Phys. Rev. D}
  {\bfseries 103} (2021) 043528}
  [\href{https://arxiv.org/abs/2011.12934}{{\ttfamily 2011.12934}}].

\bibitem{Pan:2020bur}
S.~Pan, W.~Yang and A.~Paliathanasis, \emph{{Non-linear interacting
  cosmological models after Planck 2018 legacy release and the $H_0$ tension}},
  \href{https://doi.org/10.1093/mnras/staa213}{\emph{Mon. Not. Roy. Astron.
  Soc.} {\bfseries 493} (2020) 3114}
  [\href{https://arxiv.org/abs/2002.03408}{{\ttfamily 2002.03408}}].

\bibitem{Capozziello:2020nyq}
S.~Capozziello, M.~Benetti and A.D.A.M.~Spallicci, \emph{{Addressing the
  cosmological $H_0$ tension by the Heisenberg uncertainty}},
  \href{https://doi.org/10.1007/s10701-020-00356-2}{\emph{Found. Phys.}
  {\bfseries 50} (2020) 893}
  [\href{https://arxiv.org/abs/2007.00462}{{\ttfamily 2007.00462}}].

\bibitem{Saridakis:2019qwt}
E.N.~Saridakis, S.~Myrzakul, K.~Myrzakulov and K.~Yerzhanov,
  \emph{{Cosmological applications of $F(R,T)$ gravity with dynamical curvature
  and torsion}}, \href{https://doi.org/10.1103/PhysRevD.102.023525}{\emph{Phys.
  Rev. D} {\bfseries 102} (2020) 023525}
  [\href{https://arxiv.org/abs/1912.03882}{{\ttfamily 1912.03882}}].

\bibitem{Escamilla-Rivera:2019ulu}
C.~Escamilla-Rivera and J.~Levi~Said, \emph{{Cosmological viable models in
  $f(T,B)$ theory as solutions to the $H_0$ tension}},
  \href{https://doi.org/10.1088/1361-6382/ab939c}{\emph{Class. Quant. Grav.}
  {\bfseries 37} (2020) 165002}
  [\href{https://arxiv.org/abs/1909.10328}{{\ttfamily 1909.10328}}].

\bibitem{DiValentino:2019jae}
E.~Di~Valentino, A.~Melchiorri, O.~Mena and S.~Vagnozzi, \emph{{Nonminimal dark
  sector physics and cosmological tensions}},
  \href{https://doi.org/10.1103/PhysRevD.101.063502}{\emph{Phys. Rev. D}
  {\bfseries 101} (2020) 063502}
  [\href{https://arxiv.org/abs/1910.09853}{{\ttfamily 1910.09853}}].

\bibitem{Benevento:2020fev}
G.~Benevento, W.~Hu and M.~Raveri, \emph{{Can Late Dark Energy Transitions
  Raise the Hubble constant?}},
  \href{https://doi.org/10.1103/PhysRevD.101.103517}{\emph{Phys. Rev. D}
  {\bfseries 101} (2020) 103517}
  [\href{https://arxiv.org/abs/2002.11707}{{\ttfamily 2002.11707}}].

\bibitem{Banerjee:2020xcn}
A.~Banerjee, H.~Cai, L.~Heisenberg, E.O.~Colg\'ain, M.M.~Sheikh-Jabbari and
  T.~Yang, \emph{{Hubble sinks in the low-redshift swampland}},
  \href{https://doi.org/10.1103/PhysRevD.103.L081305}{\emph{Phys. Rev. D}
  {\bfseries 103} (2021) L081305}
  [\href{https://arxiv.org/abs/2006.00244}{{\ttfamily 2006.00244}}].

\bibitem{Elizalde:2020mfs}
E.~Elizalde, M.~Khurshudyan, S.D.~Odintsov and R.~Myrzakulov, \emph{{Analysis
  of the $H_0$ tension problem in the Universe with viscous dark fluid}},
  \href{https://doi.org/10.1103/PhysRevD.102.123501}{\emph{Phys. Rev. D}
  {\bfseries 102} (2020) 123501}
  [\href{https://arxiv.org/abs/2006.01879}{{\ttfamily 2006.01879}}].

\bibitem{DeFelice:2020cpt}
A.~De~Felice, S.~Mukohyama and M.C.~Pookkillath, \emph{{Addressing $H_0$
  tension by means of VCDM}},
  \href{https://doi.org/10.1016/j.physletb.2021.136201}{\emph{Phys. Lett. B}
  {\bfseries 816} (2021) 136201}
  [\href{https://arxiv.org/abs/2009.08718}{{\ttfamily 2009.08718}}].

\bibitem{Haridasu:2020pms}
B.S.~Haridasu, M.~Viel and N.~Vittorio, \emph{{Sources of $H_0$-tension in dark
  energy scenarios}},
  \href{https://doi.org/10.1103/PhysRevD.103.063539}{\emph{Phys. Rev. D}
  {\bfseries 103} (2021) 063539}
  [\href{https://arxiv.org/abs/2012.10324}{{\ttfamily 2012.10324}}].

\bibitem{Seto:2021xua}
O.~Seto and Y.~Toda, \emph{{Comparing early dark energy and extra radiation
  solutions to the Hubble tension with BBN}},
  \href{https://doi.org/10.1103/PhysRevD.103.123501}{\emph{Phys. Rev. D}
  {\bfseries 103} (2021) 123501}
  [\href{https://arxiv.org/abs/2101.03740}{{\ttfamily 2101.03740}}].

\bibitem{Adi:2020qqf}
T.~Adi and E.D.~Kovetz, \emph{{Can conformally coupled modified gravity solve
  the Hubble tension?}},
  \href{https://doi.org/10.1103/PhysRevD.103.023530}{\emph{Phys. Rev. D}
  {\bfseries 103} (2021) 023530}
  [\href{https://arxiv.org/abs/2011.13853}{{\ttfamily 2011.13853}}].

\bibitem{Ballardini:2020iws}
M.~Ballardini, M.~Braglia, F.~Finelli, D.~Paoletti, A.A.~Starobinsky and
  C.~Umilt\`a, \emph{{Scalar-tensor theories of gravity, neutrino physics, and
  the $H_0$ tension}},
  \href{https://doi.org/10.1088/1475-7516/2020/10/044}{\emph{JCAP} {\bfseries
  10} (2020) 044} [\href{https://arxiv.org/abs/2004.14349}{{\ttfamily
  2004.14349}}].

\bibitem{LinaresCedeno:2020uxx}
F.X.~Linares Cede\~no and U.~Nucamendi, \emph{{Revisiting cosmological
  diffusion models in Unimodular Gravity and the $H_0$ tension}},
  \href{https://doi.org/10.1016/j.dark.2021.100807}{\emph{Phys. Dark Univ.}
  {\bfseries 32} (2021) 100807}
  [\href{https://arxiv.org/abs/2009.10268}{{\ttfamily 2009.10268}}].

\bibitem{Odintsov:2020qzd}
S.D.~Odintsov, D.~S\'aez-Chill\'on~G\'omez and G.S.~Sharov, \emph{{Analyzing
  the $H_0$ tension in $F(R)$ gravity models}},
  \href{https://doi.org/10.1016/j.nuclphysb.2021.115377}{\emph{Nucl. Phys. B}
  {\bfseries 966} (2021) 115377}
  [\href{https://arxiv.org/abs/2011.03957}{{\ttfamily 2011.03957}}].

\bibitem{Alestas:2021xes}
G.~Alestas and L.~Perivolaropoulos, \emph{{Late-time approaches to the Hubble
  tension deforming H(z), worsen the growth tension}},
  \href{https://doi.org/10.1093/mnras/stab1070}{\emph{Mon. Not. Roy. Astron.
  Soc.} {\bfseries 504} (2021) 3956}
  [\href{https://arxiv.org/abs/2103.04045}{{\ttfamily 2103.04045}}].

\bibitem{Elizalde:2021kmo}
E.~Elizalde, J.~Gluza and M.~Khurshudyan, \emph{{An approach to cold dark
  matter deviation and the $H_{0}$ tension problem by using machine learning}},
   \href{https://arxiv.org/abs/2104.01077}{{\ttfamily 2104.01077}}.

\bibitem{Basilakos:2023xof}
S.~Basilakos, D.V.~Nanopoulos, T.~Papanikolaou, E.N.~Saridakis and C.~Tzerefos,
  \emph{{Signatures of Superstring theory in NANOGrav}},
  \href{https://arxiv.org/abs/2307.08601}{{\ttfamily 2307.08601}}.

\bibitem{Krishnan:2021dyb}
C.~Krishnan, R.~Mohayaee, E.O.~Colg\'ain, M.M.~Sheikh-Jabbari and L.~Yin,
  \emph{{Does Hubble tension signal a breakdown in FLRW cosmology?}},
  \href{https://doi.org/10.1088/1361-6382/ac1a81}{\emph{Class. Quant. Grav.}
  {\bfseries 38} (2021) 184001}
  [\href{https://arxiv.org/abs/2105.09790}{{\ttfamily 2105.09790}}].

\bibitem{Papanikolaou:2022did}
T.~Papanikolaou, A.~Lymperis, S.~Lola and E.N.~Saridakis, \emph{{Primordial
  black holes and gravitational waves from non-canonical inflation}},
  \href{https://doi.org/10.1088/1475-7516/2023/03/003}{\emph{JCAP} {\bfseries
  03} (2023) 003} [\href{https://arxiv.org/abs/2211.14900}{{\ttfamily
  2211.14900}}].

\bibitem{Theodoropoulos:2021hkk}
A.~Theodoropoulos and L.~Perivolaropoulos, \emph{{The Hubble Tension, the M
  Crisis of Late Time H(z) Deformation Models and the Reconstruction of
  Quintessence Lagrangians}},
  \href{https://doi.org/10.3390/universe7080300}{\emph{Universe} {\bfseries 7}
  (2021) 300} [\href{https://arxiv.org/abs/2109.06256}{{\ttfamily
  2109.06256}}].

\bibitem{Papanikolaou:2023crz}
T.~Papanikolaou, \emph{{Primordial black holes in loop quantum cosmology: the
  effect on the threshold}},
  \href{https://doi.org/10.1088/1361-6382/acd97d}{\emph{Class. Quant. Grav.}
  {\bfseries 40} (2023) 134001}
  [\href{https://arxiv.org/abs/2301.11439}{{\ttfamily 2301.11439}}].

\bibitem{Abdalla:2022yfr}
E.~Abdalla et~al., \emph{{Cosmology intertwined: A review of the particle
  physics, astrophysics, and cosmology associated with the cosmological
  tensions and anomalies}},
  \href{https://doi.org/10.1016/j.jheap.2022.04.002}{\emph{JHEAp} {\bfseries
  34} (2022) 49} [\href{https://arxiv.org/abs/2203.06142}{{\ttfamily
  2203.06142}}].

\bibitem{Lymperis:2018iuz}
A.~Lymperis and E.N.~Saridakis, \emph{{Modified cosmology through nonextensive
  horizon thermodynamics}},
  \href{https://doi.org/10.1140/epjc/s10052-018-6480-y}{\emph{Eur. Phys. J. C}
  {\bfseries 78} (2018) 993}
  [\href{https://arxiv.org/abs/1806.04614}{{\ttfamily 1806.04614}}].

\bibitem{Jacobson:1995ab}
T.~Jacobson, \emph{{Thermodynamics of space-time: The Einstein equation of
  state}}, \href{https://doi.org/10.1103/PhysRevLett.75.1260}{\emph{Phys. Rev.
  Lett.} {\bfseries 75} (1995) 1260}
  [\href{https://arxiv.org/abs/gr-qc/9504004}{{\ttfamily gr-qc/9504004}}].

\bibitem{Padmanabhan:2003gd}
T.~Padmanabhan, \emph{{Gravity and the thermodynamics of horizons}},
  \href{https://doi.org/10.1016/j.physrep.2004.10.003}{\emph{Phys. Rept.}
  {\bfseries 406} (2005) 49}
  [\href{https://arxiv.org/abs/gr-qc/0311036}{{\ttfamily gr-qc/0311036}}].

\bibitem{Padmanabhan:2009vy}
T.~Padmanabhan, \emph{{Thermodynamical Aspects of Gravity: New insights}},
  \href{https://doi.org/10.1088/0034-4885/73/4/046901}{\emph{Rept. Prog. Phys.}
  {\bfseries 73} (2010) 046901}
  [\href{https://arxiv.org/abs/0911.5004}{{\ttfamily 0911.5004}}].

\bibitem{Tsallis:1987eu}
C.~Tsallis, \emph{{Possible Generalization of Boltzmann-Gibbs Statistics}},
  \href{https://doi.org/10.1007/BF01016429}{\emph{J. Statist. Phys.} {\bfseries
  52} (1988) 479}.

\bibitem{Lyra:1997ggy}
M.L.~Lyra and C.~Tsallis, \emph{{Nonextensivity and Multifractality in
  Low-Dimensional Dissipative Systems}},
  \href{https://doi.org/10.1103/PhysRevLett.80.53}{\emph{Phys. Rev. Lett.}
  {\bfseries 80} (1998) 53}
  [\href{https://arxiv.org/abs/cond-mat/9709226}{{\ttfamily
  cond-mat/9709226}}].

\bibitem{Wilk:1999dr}
G.~Wilk and Z.~Wlodarczyk, \emph{{On the interpretation of nonextensive
  parameter q in Tsallis statistics and Levy distributions}},
  \href{https://doi.org/10.1103/PhysRevLett.84.2770}{\emph{Phys. Rev. Lett.}
  {\bfseries 84} (2000) 2770}
  [\href{https://arxiv.org/abs/hep-ph/9908459}{{\ttfamily hep-ph/9908459}}].

\bibitem{Sheykhi:2018dpn}
A.~Sheykhi, \emph{{Modified Friedmann Equations from Tsallis Entropy}},
  \href{https://doi.org/10.1016/j.physletb.2018.08.036}{\emph{Phys. Lett. B}
  {\bfseries 785} (2018) 118}
  [\href{https://arxiv.org/abs/1806.03996}{{\ttfamily 1806.03996}}].

\bibitem{Lalus:2019lzq}
H.F.~Lalus and G.~Hikmawan, \emph{{Analytical solutions of modified Friedmann
  equation in Tsallis Cosmology for nonflat universe}}, {\emph{International
  Journal of Innovation, Creativity and Change} {\bfseries 5} (2019) 638}.

\bibitem{Nojiri:2019skr}
S.~Nojiri, S.D.~Odintsov and E.N.~Saridakis, \emph{{Modified cosmology from
  extended entropy with varying exponent}},
  \href{https://doi.org/10.1140/epjc/s10052-019-6740-5}{\emph{Eur. Phys. J. C}
  {\bfseries 79} (2019) 242}
  [\href{https://arxiv.org/abs/1903.03098}{{\ttfamily 1903.03098}}].

\bibitem{Geng:2019shx}
C.-Q.~Geng, Y.-T.~Hsu, J.-R.~Lu and L.~Yin, \emph{{Modified Cosmology Models
  from Thermodynamical Approach}},
  \href{https://doi.org/10.1140/epjc/s10052-019-7476-y}{\emph{Eur. Phys. J. C}
  {\bfseries 80} (2020) 21} [\href{https://arxiv.org/abs/1911.06046}{{\ttfamily
  1911.06046}}].

\bibitem{Ghoshal:2021ief}
A.~Ghoshal and G.~Lambiase, \emph{{Constraints on Tsallis Cosmology from Big
  Bang Nucleosynthesis and Dark Matter Freeze-out}},
  \href{https://arxiv.org/abs/2104.11296}{{\ttfamily 2104.11296}}.

\bibitem{Luciano:2021ndh}
G.G.~Luciano, \emph{{Tsallis statistics and generalized uncertainty
  principle}},
  \href{https://doi.org/10.1140/epjc/s10052-021-09486-x}{\emph{Eur. Phys. J. C}
  {\bfseries 81} (2021) 672}.

\bibitem{Zamora:2022cqz}
D.J.~Zamora and C.~Tsallis, \emph{{Thermodynamically consistent entropic
  late-time cosmological acceleration}},
  \href{https://doi.org/10.1140/epjc/s10052-022-10645-x}{\emph{Eur. Phys. J. C}
  {\bfseries 82} (2022) 689}
  [\href{https://arxiv.org/abs/2201.03385}{{\ttfamily 2201.03385}}].

\bibitem{Luciano:2022ely}
G.G.~Luciano and J.~Gine, \emph{{Baryogenesis in non-extensive Tsallis
  Cosmology}},
  \href{https://doi.org/10.1016/j.physletb.2022.137352}{\emph{Phys. Lett. B}
  {\bfseries 833} (2022) 137352}
  [\href{https://arxiv.org/abs/2204.02723}{{\ttfamily 2204.02723}}].

\bibitem{Nojiri:2022dkr}
S.~Nojiri, S.D.~Odintsov and T.~Paul, \emph{{Early and late universe
  holographic cosmology from a new generalized entropy}},
  \href{https://doi.org/10.1016/j.physletb.2022.137189}{\emph{Phys. Lett. B}
  {\bfseries 831} (2022) 137189}
  [\href{https://arxiv.org/abs/2205.08876}{{\ttfamily 2205.08876}}].

\bibitem{Jizba:2022bfz}
P.~Jizba and G.~Lambiase, \emph{{Tsallis cosmology and its applications in dark
  matter physics with focus on IceCube high-energy neutrino data}},
  \href{https://doi.org/10.1140/epjc/s10052-022-11113-2}{\emph{Eur. Phys. J. C}
  {\bfseries 82} (2022) 1123}
  [\href{https://arxiv.org/abs/2206.12910}{{\ttfamily 2206.12910}}].

\bibitem{Tsallis:2012js}
C.~Tsallis and L.J.L.~Cirto, \emph{{Black hole thermodynamical entropy}},
  \href{https://doi.org/10.1140/epjc/s10052-013-2487-6}{\emph{Eur. Phys. J. C}
  {\bfseries 73} (2013) 2487}
  [\href{https://arxiv.org/abs/1202.2154}{{\ttfamily 1202.2154}}].

\bibitem{Cai:2005ra}
R.-G.~Cai and S.P.~Kim, \emph{{First law of thermodynamics and Friedmann
  equations of Friedmann-Robertson-Walker universe}},
  \href{https://doi.org/10.1088/1126-6708/2005/02/050}{\emph{JHEP} {\bfseries
  02} (2005) 050} [\href{https://arxiv.org/abs/hep-th/0501055}{{\ttfamily
  hep-th/0501055}}].

\bibitem{Akbar:2006kj}
M.~Akbar and R.-G.~Cai, \emph{{Thermodynamic Behavior of Friedmann Equations at
  Apparent Horizon of FRW Universe}},
  \href{https://doi.org/10.1103/PhysRevD.75.084003}{\emph{Phys. Rev. D}
  {\bfseries 75} (2007) 084003}
  [\href{https://arxiv.org/abs/hep-th/0609128}{{\ttfamily hep-th/0609128}}].

\bibitem{Izquierdo:2005ku}
G.~Izquierdo and D.~Pavon, \emph{{Dark energy and the generalized second law}},
  \href{https://doi.org/10.1016/j.physletb.2005.12.040}{\emph{Phys. Lett. B}
  {\bfseries 633} (2006) 420}
  [\href{https://arxiv.org/abs/astro-ph/0505601}{{\ttfamily
  astro-ph/0505601}}].

\bibitem{Cai:2006rs}
R.-G.~Cai and L.-M.~Cao, \emph{{Unified first law and thermodynamics of
  apparent horizon in FRW universe}},
  \href{https://doi.org/10.1103/PhysRevD.75.064008}{\emph{Phys. Rev. D}
  {\bfseries 75} (2007) 064008}
  [\href{https://arxiv.org/abs/gr-qc/0611071}{{\ttfamily gr-qc/0611071}}].

\bibitem{Akbar:2006er}
M.~Akbar and R.-G.~Cai, \emph{{Friedmann equations of FRW universe in
  scalar-tensor gravity, f(R) gravity and first law of thermodynamics}},
  \href{https://doi.org/10.1016/j.physletb.2006.02.035}{\emph{Phys. Lett. B}
  {\bfseries 635} (2006) 7}
  [\href{https://arxiv.org/abs/hep-th/0602156}{{\ttfamily hep-th/0602156}}].

\bibitem{Paranjape:2006ca}
A.~Paranjape, S.~Sarkar and T.~Padmanabhan, \emph{{Thermodynamic route to field
  equations in Lancos-Lovelock gravity}},
  \href{https://doi.org/10.1103/PhysRevD.74.104015}{\emph{Phys. Rev. D}
  {\bfseries 74} (2006) 104015}
  [\href{https://arxiv.org/abs/hep-th/0607240}{{\ttfamily hep-th/0607240}}].

\bibitem{Sheykhi:2007zp}
A.~Sheykhi, B.~Wang and R.-G.~Cai, \emph{{Thermodynamical properties of
  apparent horizon in warped DGP braneworld}},
  \href{https://doi.org/10.1016/j.nuclphysb.2007.04.028}{\emph{Nucl. Phys. B}
  {\bfseries 779} (2007) 1}
  [\href{https://arxiv.org/abs/hep-th/0701198}{{\ttfamily hep-th/0701198}}].

\bibitem{Jamil:2009eb}
M.~Jamil, E.N.~Saridakis and M.R.~Setare, \emph{{Thermodynamics of dark energy
  interacting with dark matter and radiation}},
  \href{https://doi.org/10.1103/PhysRevD.81.023007}{\emph{Phys. Rev. D}
  {\bfseries 81} (2010) 023007}
  [\href{https://arxiv.org/abs/0910.0822}{{\ttfamily 0910.0822}}].

\bibitem{Cai:2009ph}
R.-G.~Cai and N.~Ohta, \emph{{Horizon Thermodynamics and Gravitational Field
  Equations in Horava-Lifshitz Gravity}},
  \href{https://doi.org/10.1103/PhysRevD.81.084061}{\emph{Phys. Rev. D}
  {\bfseries 81} (2010) 084061}
  [\href{https://arxiv.org/abs/0910.2307}{{\ttfamily 0910.2307}}].

\bibitem{Wang:2009zv}
M.~Wang, J.~Jing, C.~Ding and S.~Chen, \emph{{First law of thermodynamics in IR
  modified Ho\v{r}ava-Lifshitz gravity}},
  \href{https://doi.org/10.1103/PhysRevD.81.083006}{\emph{Phys. Rev. D}
  {\bfseries 81} (2010) 083006}
  [\href{https://arxiv.org/abs/0912.4832}{{\ttfamily 0912.4832}}].

\bibitem{Gim:2014nba}
Y.~Gim, W.~Kim and S.-H.~Yi, \emph{{The first law of thermodynamics in Lifshitz
  black holes revisited}},
  \href{https://doi.org/10.1007/JHEP07(2014)002}{\emph{JHEP} {\bfseries 07}
  (2014) 002} [\href{https://arxiv.org/abs/1403.4704}{{\ttfamily 1403.4704}}].

\bibitem{Fan:2014ala}
Z.-Y.~Fan and H.~Lu, \emph{{Thermodynamical First Laws of Black Holes in
  Quadratically-Extended Gravities}},
  \href{https://doi.org/10.1103/PhysRevD.91.064009}{\emph{Phys. Rev. D}
  {\bfseries 91} (2015) 064009}
  [\href{https://arxiv.org/abs/1501.00006}{{\ttfamily 1501.00006}}].

\bibitem{Saridakis:2020lrg}
E.N.~Saridakis, \emph{{Modified cosmology through spacetime thermodynamics and
  Barrow horizon entropy}},
  \href{https://doi.org/10.1088/1475-7516/2020/07/031}{\emph{JCAP} {\bfseries
  07} (2020) 031} [\href{https://arxiv.org/abs/2006.01105}{{\ttfamily
  2006.01105}}].

\bibitem{Hernandez-Almada:2021rjs}
A.~Hern\'andez-Almada, G.~Leon, J.~Maga\~na, M.A.~Garc\'\i{}a-Aspeitia,
  V.~Motta, E.N.~Saridakis et~al., \emph{{Observational constraints and
  dynamical analysis of Kaniadakis horizon-entropy cosmology}},
  \href{https://doi.org/10.1093/mnras/stac795}{\emph{Mon. Not. Roy. Astron.
  Soc.} {\bfseries 512} (2022) 5122}
  [\href{https://arxiv.org/abs/2112.04615}{{\ttfamily 2112.04615}}].

\bibitem{Sheykhi:2022gzb}
A.~Sheykhi and B.~Farsi, \emph{{Growth of perturbations in Tsallis and Barrow
  cosmology}},
  \href{https://doi.org/10.1140/epjc/s10052-022-11044-y}{\emph{Eur. Phys. J. C}
  {\bfseries 82} (2022) 1111}
  [\href{https://arxiv.org/abs/2205.04138}{{\ttfamily 2205.04138}}].

\bibitem{Heisenberg:2022lob}
L.~Heisenberg, H.~Villarrubia-Rojo and J.~Zosso, \emph{{Simultaneously solving
  the H0 and \ensuremath{\sigma}8 tensions with late dark energy}},
  \href{https://doi.org/10.1016/j.dark.2022.101163}{\emph{Phys. Dark Univ.}
  {\bfseries 39} (2023) 101163}
  [\href{https://arxiv.org/abs/2201.11623}{{\ttfamily 2201.11623}}].

\bibitem{Heisenberg:2022gqk}
L.~Heisenberg, H.~Villarrubia-Rojo and J.~Zosso, \emph{{Can late-time
  extensions solve the H0 and \ensuremath{\sigma}8 tensions?}},
  \href{https://doi.org/10.1103/PhysRevD.106.043503}{\emph{Phys. Rev. D}
  {\bfseries 106} (2022) 043503}
  [\href{https://arxiv.org/abs/2202.01202}{{\ttfamily 2202.01202}}].

\bibitem{Asghari:2021lzu}
M.~Asghari and A.~Sheykhi, \emph{{Observational constraints on Tsallis modified
  gravity}}, \href{https://doi.org/10.1093/mnras/stab2671}{\emph{Mon. Not. Roy.
  Astron. Soc.} {\bfseries 508} (2021) 2855}
  [\href{https://arxiv.org/abs/2106.15551}{{\ttfamily 2106.15551}}].

\bibitem{Yu:2017iju}
H.~Yu, B.~Ratra and F.-Y.~Wang, \emph{{Hubble Parameter and Baryon Acoustic
  Oscillation Measurement Constraints on the Hubble Constant, the Deviation
  from the Spatially Flat \ensuremath{\Lambda}CDM Model, the
  Deceleration\textendash{}Acceleration Transition Redshift, and Spatial
  Curvature}}, \href{https://doi.org/10.3847/1538-4357/aab0a2}{\emph{Astrophys.
  J.} {\bfseries 856} (2018) 3}
  [\href{https://arxiv.org/abs/1711.03437}{{\ttfamily 1711.03437}}].

\bibitem{Jimenez:2001gg}
R.~Jimenez and A.~Loeb, \emph{{Constraining cosmological parameters based on
  relative galaxy ages}},
  \href{https://doi.org/10.1086/340549}{\emph{Astrophys. J.} {\bfseries 573}
  (2002) 37} [\href{https://arxiv.org/abs/astro-ph/0106145}{{\ttfamily
  astro-ph/0106145}}].

\bibitem{Wang:2017wia}
Y.~Wang, G.-B.~Zhao, C.-H.~Chuang, M.~Pellejero-Ibanez, C.~Zhao, F.-S.~Kitaura
  et~al., \emph{{The clustering of galaxies in the completed SDSS-III Baryon
  Oscillation Spectroscopic Survey: a tomographic analysis of structure growth
  and expansion rate from anisotropic galaxy clustering}},
  \href{https://doi.org/10.1093/mnras/sty2449}{\emph{Mon. Not. Roy. Astron.
  Soc.} {\bfseries 481} (2018) 3160}
  [\href{https://arxiv.org/abs/1709.05173}{{\ttfamily 1709.05173}}].

\bibitem{Gil-Marin:2018cgo}
H.~Gil-Mar\'\i{}n et~al., \emph{{The clustering of the SDSS-IV extended Baryon
  Oscillation Spectroscopic Survey DR14 quasar sample: structure growth rate
  measurement from the anisotropic quasar power spectrum in the redshift range
  $0.8 < z < 2.2$}}, \href{https://doi.org/10.1093/mnras/sty453}{\emph{Mon.
  Not. Roy. Astron. Soc.} {\bfseries 477} (2018) 1604}
  [\href{https://arxiv.org/abs/1801.02689}{{\ttfamily 1801.02689}}].

\bibitem{Hou:2018yny}
J.~Hou et~al., \emph{{The clustering of the SDSS-IV extended Baryon Oscillation
  Spectroscopic Survey DR14 quasar sample: anisotropic clustering analysis in
  configuration-space}},
  \href{https://doi.org/10.1093/mnras/sty1984}{\emph{Mon. Not. Roy. Astron.
  Soc.} {\bfseries 480} (2018) 2521}
  [\href{https://arxiv.org/abs/1801.02656}{{\ttfamily 1801.02656}}].

\bibitem{Zhao:2018gvb}
G.-B.~Zhao et~al., \emph{{The clustering of the SDSS-IV extended Baryon
  Oscillation Spectroscopic Survey DR14 quasar sample: a tomographic
  measurement of cosmic structure growth and expansion rate based on optimal
  redshift weights}}, \href{https://doi.org/10.1093/mnras/sty2845}{\emph{Mon.
  Not. Roy. Astron. Soc.} {\bfseries 482} (2019) 3497}
  [\href{https://arxiv.org/abs/1801.03043}{{\ttfamily 1801.03043}}].

\bibitem{Cai:2009zp}
Y.-F.~Cai, E.N.~Saridakis, M.R.~Setare and J.-Q.~Xia, \emph{{Quintom Cosmology:
  Theoretical implications and observations}},
  \href{https://doi.org/10.1016/j.physrep.2010.04.001}{\emph{Phys. Rept.}
  {\bfseries 493} (2010) 1} [\href{https://arxiv.org/abs/0909.2776}{{\ttfamily
  0909.2776}}].

\bibitem{Barrow:2020kug}
J.D.~Barrow, S.~Basilakos and E.N.~Saridakis, \emph{{Big Bang Nucleosynthesis
  constraints on Barrow entropy}},
  \href{https://doi.org/10.1016/j.physletb.2021.136134}{\emph{Phys. Lett. B}
  {\bfseries 815} (2021) 136134}
  [\href{https://arxiv.org/abs/2010.00986}{{\ttfamily 2010.00986}}].

\bibitem{Lymperis:2021qty}
A.~Lymperis, S.~Basilakos and E.N.~Saridakis, \emph{{Modified cosmology through
  Kaniadakis horizon entropy}},
  \href{https://doi.org/10.1140/epjc/s10052-021-09852-9}{\emph{Eur. Phys. J. C}
  {\bfseries 81} (2021) 1037}
  [\href{https://arxiv.org/abs/2108.12366}{{\ttfamily 2108.12366}}].

\end{thebibliography}\endgroup

\end{document}